# Space Weather Observations, Modeling, and Alerts in Support of Human Exploration of Mars

**James L. Green[1*], Chuanfei Dong[2], Michael Hesse[3], C. Alex Young[4], Vladimir Airapetian[4,5]**

[1]NASA Headquarters, Office of the Chief Scientist, Washington D.C., USA

[2]Princeton Plasma Physics Laboratory and Department of Astrophysical Sciences, Princeton University, Princeton, NJ, USA

[3]NASA Ames Research Center, Mountain View, CA, USA

[4]NASA Goddard Space Flight Center, Greenbelt, MD, USA

[5]American University, Washington, DC, USA

**\*Correspondence:** james.green@nasa.gov

## Abstract

Space weather observations and modeling at Mars have begun but they must be significantly increased to support the future of Human Exploration on the Red Planet. A comprehensive space weather understanding of a planet without a global magnetosphere and a thin atmosphere is very different from our situation at Earth so there is substantial fundamental research remaining. It is expected that the development of suitable models will lead to a comprehensive operational Mars space weather alert (MSWA) system that would provide rapid dissemination of information to Earth controllers, astronauts in transit, and those in the exploration zone (EZ) on the surface by producing alerts that are delivered rapidly and are actionable. To illustrate the importance of such a system, we use a magnetohydrodynamic code to model an extreme Carrington-type coronal mass ejection (CME) event at Mars. The results show a significant induced surface field of nearly 3000 nT on the dayside that could radically affect unprotected electrical systems that would dramatically impact human survival on Mars. Other associated problems include coronal mass ejection (CME) shock-driven acceleration of solar energetic particles producing large doses of ionizing radiation at the Martian surface. In summary, along with working more closely with international partners, the next Heliophysics Decadal Survey must include a new initiative to meet expected demands for space weather forecasting in support of humans living and working on the surface of Mars. It will require significant effort to coordinate NASA and the international community contributions.

# 1    Introduction

Within the Science Mission Directorate (SMD) at NASA the Planetary Science Division and the Heliophysics Division collaborate in several areas such as understanding how the solar wind interacts with planets that have magnetospheres and those that do not currently have magnetic fields. From a planetary science perspective, data in these areas are used to understand the evolution of solar system bodies including planets, moons, and small bodies. The Heliophysics discipline is the "study of the nature of the Sun, and how it influences the very nature of space and, in turn, the atmospheres of planets" (https://science.nasa.gov/heliophysics). In addition, Heliophysics also provides societal benefits involving understanding and predicting the effects of space weather on our society. At Earth, the Heliophysics Science Division at NASA Goddard Space Flight Center has developed the Community Coordinated Modeling Center (CCMC), which supports and performs research in space weather modeling, through a partnership between eight agencies (including NASA, National Oceanic and Atmospheric Administration, and National Science Foundation) and numerous modeling groups across different agencies and institutions. Therefore, the bulk of the past heliophysics research has concentrated on the Sun-Earth environment providing little support for space weather research at Mars in comparison.

The Planetary Science Division has made a significant investment in modeling the atmosphere of Mars up to about 100 km through the development of the Mars Climate Modeling Center (MCMC) and is getting ready to support human exploration of the Red Planet (see Chapter 19 of NAS, 2022). Data from the Planetary Science Division's orbiters and landers are being assimilated into the MCMC atmospheric models on a regular basis enabling a detailed understanding of the global circulation of the atmosphere (temperature, pressure, and wind velocity). Recently, dust storm dynamics have also been added to the capability of the MCMC. Due to the very thin atmosphere and maintaining only an induced magnetosphere, in addition to the cosmic rays, space weather effects extend all the way to the surface of Mars making the Red Planet even more hazardous to humans than the Earth.

Today NASA's human exploration is on the verge of returning to the Moon and then onto Mars. Within this next decade, NASA's plans to send humans to Mars will become firm with human missions going to the Red Planet in the late 2030s while some companies in the commercial sector stating their desire to arrive earlier. Therefore, the cooperation and support from the Science Mission Directorate (SMD), Space Operations Mission Directorate, and the Exploration Systems Development Directorate are needed in this decade to support this new human endeavor. Through bilateral and multi-lateral meetings SMD management must also include the space weather activities of other nations (observations and modeling), to the extent possible, to make the MSWA system a reality.

Currently, there is no coordinated effort at NASA nor with the international space research community to provide space weather modeling and prediction at Mars in a comprehensive manner. This paper advocates for such a comprehensive coordinated research and modeling effort, shows the need for such modeling, and requests that the new Heliophysics decadal study includes this approach as an essential next step in getting ready for supporting human exploration of Mars. Since space weather is in the domain of NASA's Heliophysics Division it must lead the effort and obtain a new initiative recommendation in its decadal.

# 2    Heliophysics Measurements at Mars

Research into understanding the space weather environment at Mars is already in progress (e.g., Barabash et al., 2007; Dubinin et al., 2009; Jakosky et al, 2015; Bougher et al., 2015; Dong et al., 2015a; Curry et al., 2015; Luhmann et al., 2017; Ramstad et al., 2017; Ma et al., 2018; Fang et al., 2019) but a significant amount of work still needs to be done. Starting with Mars Global Surveyor, a new set of particle and field instruments have been or will soon arrive at Mars on a variety of international missions (see Table 1). It is important to note that the missions in Table 1, taken all together, are not performing all the measurements that will be needed to develop an MSWA system. For instance, there is no continuous solar wind monitor at the Mars Lagrangian point L1 which will eventually be needed. Although missions like MAVEN would be able to make solar wind observations, that only occurs when the spacecraft is outside the Martian bow shock while an upstream monitor will always be in the solar wind. The missions in Table 1 have been funded by NASA's Planetary Science Division or from other international space agencies with the notable exception of the upcoming EscaPADE mission which is being funded by the Heliophysics Division with a new target launch readiness date of October 2024. One of the important elements of planning for the next decade of the Heliophysics Division missions should be to enhance the cooperation between the Planetary Science Division and the Heliophysics Division for joint missions going to Mars or using Mars as a gravity assist enhancing the space weather mission assets at Mars. For example, the Planetary Science Division's mission Psyche was originally planned to fly first to Mars, drop off the Heliophysics Division's mission EscaPADE, and then go on into the asteroid belt. Unfortunately, that plan changed when Psyche missed its launch window. New plans for EscaPADE are currently in formulation. Before Psyche, a missed opportunity occurred with the Planetary Science Division's mission Dawn getting a gravity assist at Mars without a secondary payload that could have been delivered to the Red Planet.

## 3    Status of Space Situational Awareness at the Moon and Mars

To take human steps on Mars, NASA is first beginning its journey with its return to the Moon starting with the Artemis program. The Moon to Mars (M2M) Space Weather Analysis Office was recently established at Goddard Space Flight Center to begin the process to characterize the space radiation environment. M2M also supports NASA robotic missions with space weather assessments and anomaly analysis support in the near-Earth environment including the Moon. M2M, in close collaboration with the CCMC and the NASA Johnson Space Center Radiation Analysis Group, is tasked with supporting the Artemis needs for space weather environment modeling, communications of radiation risks to crew health and safety, and space weather real-time analysis support. M2M with the CCMC provides both real-time space weather alerts and weekly summaries in support of robotic missions throughout the solar system.

The Space Communications and Navigation (SCaN) program office is leading the development of the operational communications system, Lunanet. The real-time space weather support from M2M is provided to Artemis through Lunanet. Lunanet is being designed to extend network services to the lunar neighborhood with the goal of ultimately going further onto Mars.

The M2M is the only organization that provides real-time space weather analysis beyond the Sun-Earth direction, but it is important to recognize that its applicability is far more comprehensive for the Moon than for Mars. The complication at Mars is due to its changing location in the solar system relative to the Earth, its induced magnetosphere, ionosphere, and atmosphere all of which need significant research, modeling development, and integration. M2M capability is an excellent starting point to be fully extended to Mars.

It is important to note that the difference in the Earth and Mars interactions with the solar wind is enormous and therefore it is expected that many new and different types of measurements will be needed at Mars that are not necessary at Earth. This is one of the main reasons why a new alert system must be developed for Mars. The MSWA should be designed to provide rapid situational awareness information to Earth controllers, astronauts in transit, and those in the exploration zone (EZ) on the surface. This means that a system needs to be developed to rapidly assess the environment producing actionable alerts for the chain of stakeholders. To accomplish this the MSWA system must maintain two basic capabilities. Performing comprehensive research and coordinated modeling of the space weather environment at Mars is the new component. The second component of the system must be used during the transit phase with astronauts going to Mars or returning to Earth from Mars, since in addition to cosmic rays, solar energetic particles (SEPs) induced hazards can become mission critical (e.g., Guo et al. 2015). For this element of the MSWA, a significant and sufficient amount of modeling has been completed and is operational in the CCMC and would be implemented.

## 4    Human Exploration at Mars

NASA's strategy for Human Exploration of the surface of Mars starts with defining an EZ. Within the ~200 km diameter, EZ is the landing sites, habitat, science sites, and *in-situ* resource utilization (ISRU) sites that will be in use for decades. Humans have had a major hand in terraforming the Earth and when they arrive at Mars their influence will begin that process in earnest on the Red Planet. Over time, there may be slow or rapid changes in the planet's atmosphere simply due to the process of human living, extracting natural resources, manufacturing, creating, consuming food, and generating green-house gasses. These changes must be measured, modeled, and understood.

The average surface pressure on Mars is about 6 mbar but if the surface pressure increases to a value above 6 mbar and the surface temperature reaches a value just a bit beyond 0 $^o$C, then liquid water could exist on its surface based on the phase diagram of water. What effect these atmospheric/surface changes will have on our physical understanding of that planetary environment remains to be explored. With cosmic rays and SEPs reaching the surface (Guo et al. 2021) and space weather variations affecting technologies such as communications, power systems, and creating strong ionospheric currents in ways unknown at the present for Mars, practical solutions can only be implemented once significant research has been completed in each of those areas and a basic understanding is obtained. It is important to note that although this paper emphasizes what role Heliophysics can play in the support of human exploration of Mars, disciplines such as planetary protection and planetary science must also make significant contributions. A coordinated effort by all relevant science disciplines and the international space science community is not only needed but is essential if the success of human exploration of the Red Planet is to be achieved.

## 5    Extremes in Space Weather at Mars

Significant computational simulation codes have been developed and are being implemented in the CCMC focused on modeling the space weather of the near Earth's magnetospheric environment including the ionosphere. These codes are available for community use and adaptation. One such adapted use is given in Fig. 1 in which we have taken the 3D BATS-R-US Mars multi-species magnetohydrodynamic (MS-MHD) code (Ma et al., 2004) and included the Martian conducting core to model the response of Mars to an extreme Carrington-type space weather event. The MS-MHD model solves a separate continuity equation for each ion species, whilst solving one momentum and one energy equation for the four ion fluids, $H^+$, $O^+$, $O_2^+$, and $CO_2^+$. The MS-MHD model includes ionospheric photochemical processes such as photoionization, charge exchange, and electron impact ionization. For this study, we extend the model lower boundary into the planetary interior by including an electrically conductive core and a resistive mantle, therefore, the newly developed Conducting-Core-Surface-to-Interplanetary-Space MS-MHD (CCSIS-MS-MHD) model solves the planetary interior, planetary ionosphere, and solar wind-Mars interaction in a self-consistent way (Dong et al., 2019). The input parameters of this extreme event are adopted from Ngwira et al. (2014) (also see Table 3 of Dong et al., 2017). Specifically, at the peak of this event, the solar wind density reaches around 425 cm$^{-3}$, and the solar wind speed and southward interplanetary magnetic field (IMF) strength reach about 2000 km/s and 200 nT. Certain assumptions (e.g., the Martian conducting core size, mantle resistivity, and fully unmagnetized Mars) were made to produce these first results that clearly show the huge surface magnetic fields during post onset. This is expected to be a significant hazard to ground electrical equipment but much more work needs to be accomplished.

What we have found out is that Mars responds to an extreme Carrington-type space weather event very differently than Earth does. The two panels of Fig. 1 show the snapshots at pre-storm and peak-storm phases. The black-filled circle represents the conducting core of Mars, and the red circle indicates the Martian surface. The white streamlines denote the magnetic fields (**B**) that can penetrate the planetary body. The color contours describe the strength of **B** in nT. Due to the induction response, the eddy currents at the conducting core surface produce a weak dipole-like field structure on the nightside that may be measured by surface magnetometers. The results from Fig. 1 show that a significant induced surface field of nearly 3000 nT on the dayside arises during the peak of the event. Such a large surface magnetic field, which will vary with rapidly changing solar wind conditions, could radically affect unprotected electrical systems that would dramatically impact human survival on Mars.

Fast energetic CMEs, a fundamental feature in the solar wind of a Carrington-type space weather event, usually drive strong and extended shocks, which are the sites of efficient particle acceleration at the shock front via the diffusive shock acceleration. The particles statistically gain energy as they cross the shock front scattering on magnetic inhomogeneities via first-order Fermi acceleration. The hardness of spectra and the maximum energy of precipitating SEPs specify the dosage of ionizing radiation reaching the Martian surface (Jolitz et al., 2017; Lingam et al., 2018), and thus can create dangers for human survival on Mars at the current epoch. Thus, the physical processes that drive CMEs and associated SEPs need to be realistically modeled by coupling 3D MHD models with kinetic models of particle acceleration and transport in the heliosphere. Fig. 2 shows the results of the recent attempt to model the 2012 May 17 SEP event by simulating the initiation of an energetic CME in the solar corona and the formation of the CME-driven shock in the lower corona (within ~5 $R_\odot$) using the CCMC-hosted coupled AWSoM and iPATH models (Li et al. 2021). Such models should be capable of capturing CME driven shocks that will provide accurate predictions of extreme SEP energy spectra and their maximum energy. This knowledge is critical in assessing the dosage of ionizing radiation at the Martian surface during severe space weather events as the crucial factor of forming long-term settlements on Mars (Da Pieve et al. 2021; Yamashiki et al. 2019).

The comprehensive MSWA system must be able to use a variety of existing models. Many of these models are currently interconnected and in operational use at CCMC (e.g., solar corona and heliosphere). Many research models (e.g., Kallio et al., 2011; Dong et al, 2014, 2015b, 2018a, 2018b; Leblanc et al., 2017; Egan et al., 2018) for the Martian induced magnetosphere, ionosphere-thermosphere-exosphere, and those in the MCMC (Haberle et al., 2019; Kahre et al., 2022) are available but are currently not integrated together. A complete gap analysis of existing models with those under development and newly developed ones will have to be made as shown in Fig.3. This type of analysis and model integration is an essential first step of a new initiative.

## 6        The Next Decadal Missions

In support of the development of the Planetary decadal survey, there were 37 white papers submitted about Mars science mission concepts to the National Academy of Sciences. There were several of these mission concepts that clearly show a need for observing Mars from a variety of orbits ranging from low altitudes, aerostationary, and out to Mars L1. These mission concepts were trying to fill critical science gaps ranging from understanding solar activity effects on Mars' system, the coupled solar wind-induced magnetosphere, ionospheric and upper atmosphere dynamics to exploring the interior structure through induced magnetosphere-interior coupling during extreme events (e.g., induction effects as shown in Fig. 1). It is important to note that the Heliophysics mission concepts discussed in these white papers were not included in the Planetary Decadal Survey (NAS, 2022). It will be up to the Heliophysics decadal to recognize the importance of these missions and prioritize them appropriately.

Another significant collaboration opportunity between human exploration and the Planetary Science Division will be on the Mars Life Explorer medium-class mission prioritized by the Planetary Decadal Survey (see Chapter 22 of NAS, 2022). This collaboration, however, is confined to better understanding Mars surface elements, such as the effect of dust, perchlorates, and other oxidizing agents in the Martian soils. To better understand the risk and mitigate the effects on humans of solar and galactic radiation, the Heliophysics Division should be responsible for providing key Heliophysics payloads. The required planning for this collaboration must begin now with the upcoming Heliophysics decadal survey.

## 7        Futuristic View of 2050

A comprehensive MSWA system will be operational. The system starts at the Sun, with $360^{\circ}$ coverage by observations and models that ingest relevant data as they become available, and that can evolve in response to new scientific insights. Many of the models used have been developed and are in active use in both the CCMC and the MSWA along with the newly developed Mars models for the induced magnetopause, ionosphere/thermosphere/exosphere, interfacing with the models in the MCMC, all of which are integrated into a seamlessly coordinated system (see Fig. 3). This whole system then provides the breadth of relevant space weather information as well as output suitable for scientific research, validation, and verification. The MSWA also permits the modeling of events based on historical data, as well as input data generated to predict future events. The MSWA system is designed to provide rapid dissemination of information to Earth controllers, astronauts in transit, and those in the EZ by producing alerts that are delivered rapidly and are actionable.

As led by NASA, Human Exploration of Mars will be both an international space agency sector and a commercial endeavor. All sectors provide significant resources and capabilities in a highly coordinated fashion. A series of key measurements are fed into the MSWA that include: 1) next-generation STEREO solar observations; 2) Mars L1 solar and solar wind observations; 3) low-altitude planet-asynchronous orbiter observations monitoring dynamical phenomena rapidly evolving in space and time; and 4) planet-wide array of ground-based magnetometers, seismic, and other meteorological instruments. In addition, the communication infrastructure also supports space weather observations. At least three spacecraft at aerostationary orbit perform the following functions: 1) relay all surface voice/video/data back to Earth; 2) provide Earth to the EZ relay capability; 3) atmospheric weather observations whose telemetry is relayed to the EZ and to Earth; 4) particles and field sensors whose telemetry is relayed to the EZ and to Earth. It should be noted that this system could also provide significant benefits for other missions in interplanetary space, or for robotic endeavors targeting scientific research of other planets.

## 8    Conclusions

Space scientists are making significant progress in understanding space weather dangers in near-Earth space, but we lack essential and fundamental knowledge in the Martian environment. As discussed in this paper, space weather observations and modeling at Mars have begun but it is largely incomplete and the solar wind, ionosphere, atmosphere, and surface features are disconnected. The Heliophysics Division must support Agency goals at Mars and that need will continue to increase over time. The Planetary Science Division has made significant contributions to supporting the Heliophysics Division science at Mars, but it cannot do this in a sustainable or significant way beyond what it has already accomplished. Since astronaut health and safety will be paramount, the Heliophysics Division has an obligation to increase its program content to the study of space weather to provide protection for travelers to Mars and residents on Martian soil, we therefore need to establish a program of research, based on modeling and interpretation of *in-situ* data from an array of current and future missions.

The choice then, is what needs to be done next and how the program will evolve to provide comprehensive space weather support for Mars human exploration activities. Therefore, the authors believe that we must first develop a comprehensive Mars space weather science/schedule roadmap to be executed delineating the gaps in research activities as a strong recommendation in the next Heliophysics Decadal. The Heliophysics Decadal must also have potential new space weather missions that are not being enabled in any other way but must be executed. Additional elements of that roadmap should include key heliophysics payloads that could be provided to the Planetary Science Division's Mars missions (orbiters, landers, rovers) through joint solicitation or strategic agreements. This approach should also apply to international space agencies and new commercial space partners as well. In other words, the Heliophysics Division needs additional funding, and it will need a ***new initiative*** to accomplish this new science.

### Author Contributions

The writing of the paper was led by JLG with equal contributions by all other authors. The 3D magnetohydrodynamic simulations were performed by CD.

### Funding

CD was supported by Princeton Plasma Physics Laboratory through the Laboratory Directed Research and Development (LDRD) Program under DOE Prime Contract No. DE-AC02-09CH11466 and by NASA through the Solar System Workings (SSW) program. VA was supported by the SEEC (Sellers Exoplanetary Environments Collaboration), which is the NASA Planetary Science Division's Internal Scientist Funding Model (ISFM).

### Acknowledgements

Resources for this work were provided by the NASA High-End Computing (HEC) Program through the NASA Advanced Supercomputing (NAS) Division at Ames Research Center. The Space Weather Modeling Framework (SWMF) that contains the BATS-R-US codes used in this study is publicly available at https://github.com/MSTEM-QUDA/SWMF. SWMF runs can also be requested via the Community Coordinated Modeling Center (CCMC) at the NASA Goddard Space Flight Center through a user-friendly web interface. For distribution of model results in this study, please contact CD.

**Table 1:** Past, Current, and near-term Mars missions with space weather relevant instruments.

| Mission | Type | Key Instrument | Key Discoveries or Expected Results |
|---|---|---|---|
| Mars Global Surveyor | Orbiter | Magnetometer | Measured significant remnant field or mini-bubble magnetospheres emanating from the surface |
| Mars Express (ESA) | Orbiter | Plasma & Radio Sounding | First comprehensive measurements of the plasma environment; ionosphere structure and dynamics |
| Curiosity | Rover | High Energy Radiation | Measured the type and amount of harmful radiation traveling to Mars and on the surface |
| MAVEN | Orbiter | Entire Payload | Solar wind-Mars interaction and the associated atmospheric loss processes; many results |
| Mars Orbiter Mission (ISRO) | Orbiter | Exosphere Neutral Composition | Thermosphere heating observed lofting neutrals to very high altitudes |
| InSight | Lander | Magnetometer | Fluctuating surface magnetic fields from interior and ionospheric currents |
| HOPE (UAE) | Orbiter | FUV Imager | Global thermosphere variability – hydrogen and oxygen coronae |
| Tianwen-1 (CNSA) | Orbiter | Plasma, Energetic Particles, | Solar wind-Mars interactions and the associated atmospheric loss processes; |

| | | Magnetometer & Sounding Radar | remnant field measurements; ionospheric structure |
|---|---|---|---|
| Zhurong (CNSA) | Lander Rover | Magnetometer | Measure surface magnetic fields (expect fluctuations and determine the source) |
| *ExoMars (ESA) | Lander Rover | Magnetometer | Measure surface magnetic fields (expect fluctuations and determine the source) |
| +EscaPADE | Twin Orbiters | Entire payload | Structures of induced magnetosphere and how it guides ion flows and atmospheric loss processes |

*Mission delayed, launch TBD; +A new target launch readiness date of October 2024.

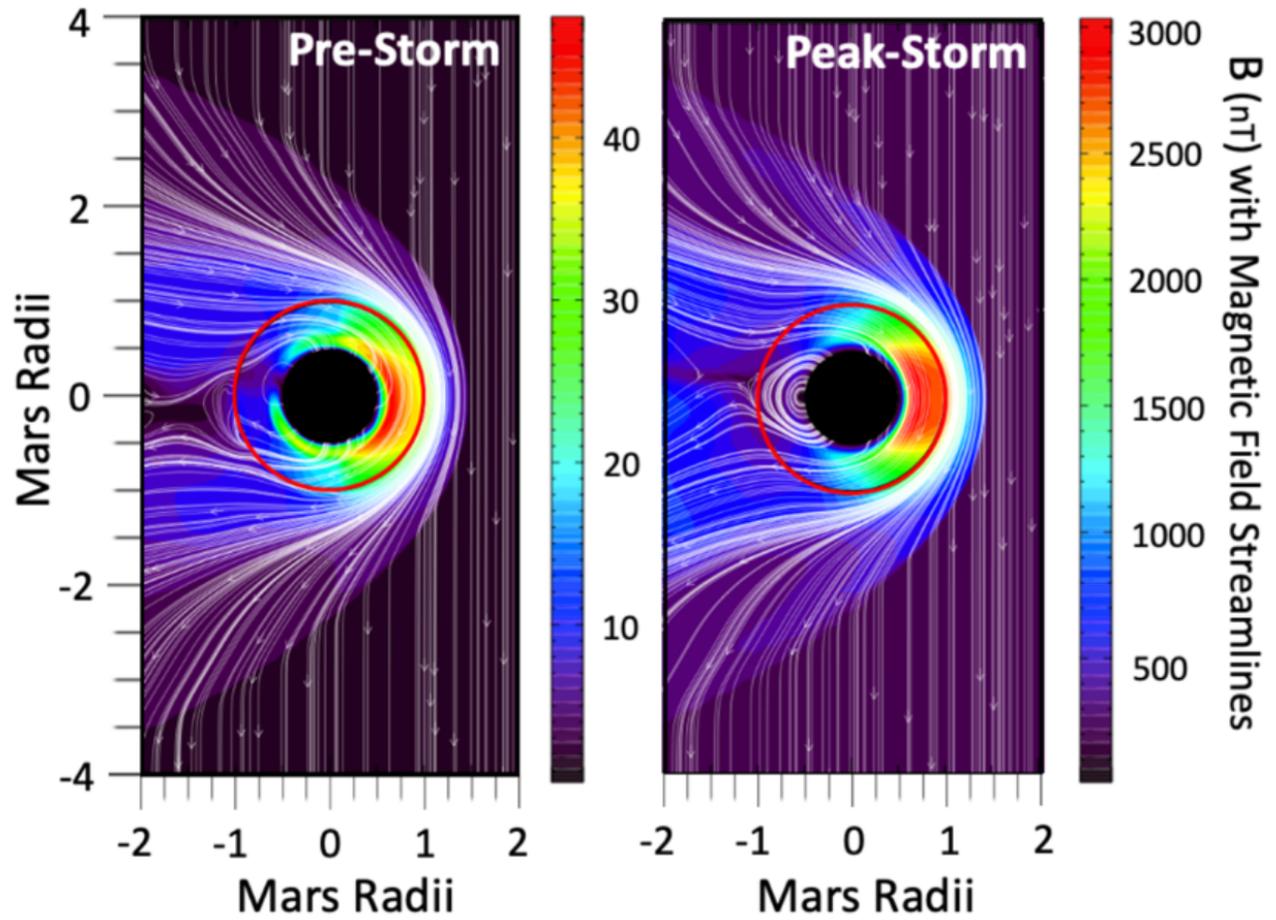

Figure 1: The modeled Martian responses to an extreme Carrington-type SW event before (**A**) and at the peak of the solar storm (**B**).

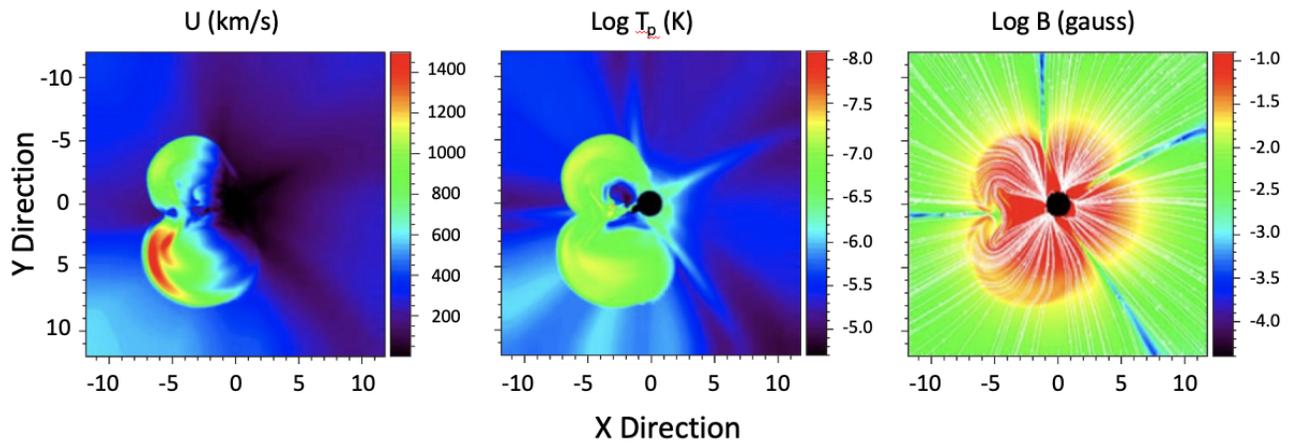

Figure 2. Z = 0 slices in the Carrington coordinates of the AWSoM MHD solution showing the radial velocity (**A**), proton temperature (**B**), and magnetic field strength (**C**) with projected field lines overlaid at t = 30 minutes after the CME initiation (adopted from Li et al., 2021)

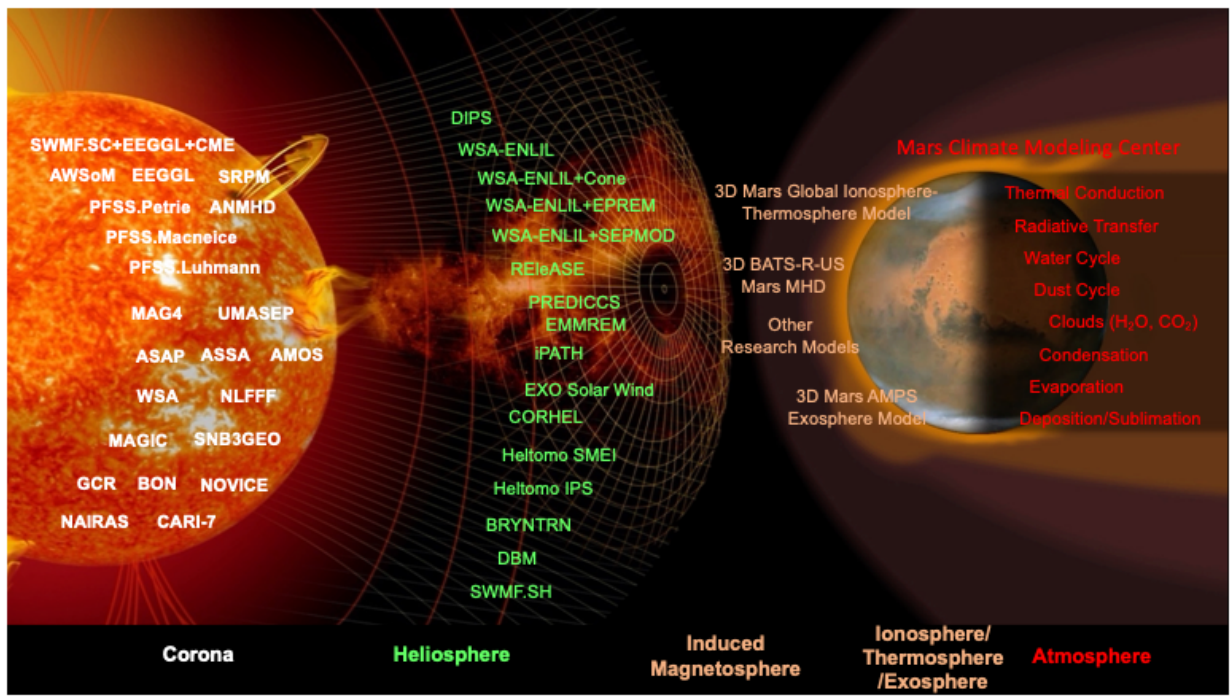

Figure 3. The comprehensive MSWA system will be using a variety of existing models, currently in operational use in the CCMC (e.g., solar corona and heliosphere), a series of new models for the induced magnetosphere, ionosphere/thermosphere/exosphere, and those in the MCMC.